\documentclass[prl, twocolumn, showpacs, nofootinbib, amsmath,amssymb, floatfix, eqsecnum]{revtex4-1}
\pdfoutput=1
\usepackage{amsmath}
\usepackage{amssymb}
\usepackage{amsthm}
\usepackage{amsfonts}
\usepackage{comment}

\usepackage{graphicx}
\usepackage{color,framed}
\usepackage{hyperref}
\usepackage{times}
\usepackage{enumerate}
\usepackage{lipsum}
\usepackage{slashed}
\usepackage{url}
\usepackage{bbm}
\usepackage{chngcntr}
\counterwithout{equation}{section}

\hypersetup{
    colorlinks=true, 
    linktoc=all,     
    linkcolor=blue,  
}

\def \beq {\begin{equation}}
\def \eeq {\end{equation}}
\def \beqa {\begin{eqnarray}}
\def \eeqa {\end{eqnarray}}
\def \bseq {\begin{subequations}}
\def \eseq {\end{subequations}}

\begin{document}

\title{Exactly solvable model for a deconfined quantum critical point in 1D}

\author{Carolyn Zhang}
\author{Michael Levin}
\affiliation{Department of Physics, Kadanoff Center for Theoretical Physics, University of Chicago, Chicago, Illinois 60637,  USA}
\date{\today}

\begin{abstract}
We construct an exactly solvable lattice model for a deconfined quantum critical point (DQCP) in (1+1) dimensions. This DQCP occurs in an unusual setting, namely at the edge of a (2+1) dimensional bosonic symmetry protected topological phase (SPT) with $\mathbb{Z}_2\times\mathbb{Z}_2$ symmetry. The DQCP describes a transition between two gapped edges that break different $\mathbb{Z}_2$ subgroups of the full $\mathbb{Z}_2\times\mathbb{Z}_2$ symmetry. Our construction is based on an exact mapping between the SPT edge theory and a $\mathbb{Z}_4$ spin chain. This mapping reveals that DQCPs in this system are directly related to ordinary $\mathbb{Z}_4$ symmetry breaking critical points. 

\end{abstract}

\maketitle

\textbf{\emph{Introduction.}}---Deconfined quantum critical points (DQCPs) describe unusual ``Landau forbidden'' phase transitions in which the unbroken symmetry group of one phase is not a subgroup of the unbroken symmetry group of the other phase\cite{senthil2004,senthil2004_2}. 
The paradigm of this kind of critical point is the hypothesized (2+1) dimensional DQCP between the valence bond solid (VBS) phase and the N\'{e}el phase on a square lattice. The VBS phase has internal $SO(3)$ rotation symmetry but spontaneously breaks $C_4$ lattice rotation symmetry, while the N\'{e}el phase has $C_4$ symmetry but breaks $SO(3)$ symmetry. Crucially, the two symmetries are intertwined: vortices of the $C_4$ symmetry carry uncompensated spin-1/2 moments\cite{levin2004}. As a result, disordering with respect to the $C_4$ symmetry can cause ordering under the $SO(3)$ symmetry, resulting in a hypothesized direct transition between the two phases. 

Thus far, DQCPs have been studied primarily using field theory and numerical methods\cite{sandvik2007,melko2008}. One reason for this is the lack of analytically tractable lattice models for DQCPs. In this work, we take a step towards a more analytical microscopic approach, by constructing an \emph{exactly solvable} lattice model for a (1+1) dimensional DQCP. The exact solvability of our model makes explicit the mechanism for the DQCP, which lies in the unusual structure of the domain walls.
This DQCP has a similar field theory description to the (1+1) dimensional DQCP that was analyzed in Refs.~\onlinecite{jiang2019,roberts2019,huang2019} using bosonization (see also Ref.~\onlinecite{mudry2019,zheng2022}). However, our DQCP involves a different lattice realization with different (non-spatial) symmetries. 

The key idea behind our solvable lattice model is to consider a DQCP in an unusual setting -- namely, at the \emph{edge} of a (2+1) dimensional symmetry protected topological (SPT) phase. SPT edge theories provide a natural setting for DQCPs because they also have intertwined symmetries\cite{chen2013,else2014}. In particular, an SPT with a ``mixed anomaly'' between two symmetries has an edge theory where domain walls of one symmetry carry fractional charge of the other symmetry\cite{qi2008,zaletel2014,wang2015spt}. Like in the VBS/N\'{e}el system, disordering with respect to one symmetry, by proliferating domain walls of that symmetry, may cause ordering with respect to the other symmetry, thereby realizing a DQCP.

We consider the simplest example of such an SPT edge theory: the edge theory of a 2D $\mathbb{Z}_2\times\mathbb{Z}_2$ symmetric bosonic SPT with a mixed anomaly between the two $\mathbb{Z}_2$ symmetries. Using an exact mapping between the SPT edge theory and a $\mathbb{Z}_4$ spin chain, we  rigorously establish the existence of a DQCP and derive the full critical theory. 

\begin{figure}[tb]
   \centering
   \includegraphics[width=.75\columnwidth]{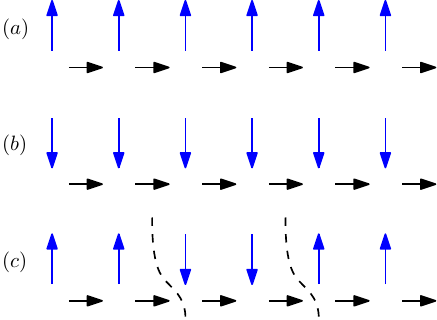} 
   \caption{(a)-(b) The two degenerate ground states of the Hamiltonian (\ref{Hssb}) that spontaneously breaks $\mathbb{Z}_{2a}$. The blue arrows represent the $\sigma_j$ spins and the black arrows represent the $\tau_{j+1/2}$ spins. Both states are eigenstates of $U_b$ with eigenvalue $+1$. (c) Domain walls occur at the boundaries between these states. A state with two $\mathbb{Z}_{2a}$ domain walls (indicated by the dashed lines) has eigenvalue $-1$ under $U_b$, meaning two $\mathbb{Z}_{2a}$ domain walls fuse to a $\mathbb{Z}_{2b}$ charge.}
   \label{fig:fig1}
   \end{figure}
   
\textbf{\emph{$\mathbb{Z}_{2a}\times\mathbb{Z}_{2b}$ SPT edge theory.}}---Our model for the SPT edge theory consists of a chain of spin-1/2's with two spins $\sigma_j$ and $\tau_{j+1/2}$ in each unit cell, labeled by $j$. The two $\mathbb{Z}_2$ symmetries, denoted by $\mathbb{Z}_{2a}$ and $\mathbb{Z}_{2b}$, are generated by unitary operators $U_a$ and $U_b$ with
\begin{equation}\label{sigmatausymm}
U_a=\prod_j\sigma^x_j\qquad U_b=\prod_j \tau^x_{j+1/2}i^{\frac{1-\sigma^z_j\sigma^z_{j+1}}{2}}.
\end{equation}
Note that $U_b$ does not act ``on-site'' in this representation: this is allowed since (\ref{sigmatausymm}) describes the effective action of the symmetries on the \emph{edge} degrees of freedom; in the original 2D spin system that describes the bulk SPT phase, both symmetries act onsite. 
   
The above symmetry action (\ref{sigmatausymm}) carries a mixed anomaly between the two symmetries. One manifestation of this mixed anomaly is that a pair of $\mathbb{Z}_{2a}$ domain walls is charged under $\mathbb{Z}_{2b}$. To see this, consider the Hamiltonian 
\begin{align}
H=-\sum_j\sigma^z_j\sigma^z_{j+1}-\sum_j\tau^x_{j+1/2}.
\label{Hssb}
\end{align}
The two degenerate ground states of this Hamiltonian, which are illustrated in Fig.~\ref{fig:fig1}(a)-(b), spontaneously break $\mathbb{Z}_{2a}$. Now consider a state with two domain walls $|\psi_{2\mathrm{dw}}\rangle$, as shown in Fig.~\ref{fig:fig1}(c). From (\ref{sigmatausymm}), we can see that such a state is actually charged under $\mathbb{Z}_{2b}$: $U_b|\psi_{2\mathrm{dw}}\rangle=-|\psi_{2\mathrm{dw}}\rangle$. Evidently, two $\mathbb{Z}_{2a}$ domain walls carry a $\mathbb{Z}_{2b}$ charge, so each domain wall can be associated with ``half" a $\mathbb{Z}_{2b}$ charge. 

\begin{figure}[tb]
   \centering
   \includegraphics[width=\columnwidth]{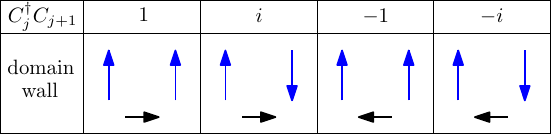} 
   \caption{A mapping between the four kinds of domain walls in the SPT edge theory and the four kinds of domain walls in the $\mathbb{Z}_4$ spin chain, which are labeled by their eigenvalues $\{1,i,-1,-i\}$ under $C_j^\dagger C_{j+1}$. As discussed in the main text, two $\mathbb{Z}_{2a}$ domain walls (second configuration) fuse to a $U_b$ charge, which is equivalent to a $\tau_{j+1/2}$ spin flip (third configuration).}
   \label{fig:fig2}
   \end{figure}

Another way to think about this anomaly is in terms of the fusion rules for domain walls. There are actually four kinds of domain walls for this system if we distinguish between states that carry different quantum numbers under the unbroken $\mathbb{Z}_{2b}$ symmetry. These four kinds of domain walls are shown in Fig.~\ref{fig:fig2}: (1) a ``no-domain wall'' state; (2) a (bare) $\mathbb{Z}_{2a}$ domain wall; (3) a $\mathbb{Z}_{2b}$ charge; (4) a composite of a $\mathbb{Z}_{2a}$ domain wall and a $\mathbb{Z}_{2b}$ charge. The fact that two $\mathbb{Z}_{2a}$ domain walls fuse to a $\mathbb{Z}_{2b}$ charge means that the fusion rules for the domain walls have a $\mathbb{Z}_4$ group structure rather than the usual $\mathbb{Z}_2\times\mathbb{Z}_2$ structure. This $\mathbb{Z}_4$ fusion structure points to a connection between our edge theory with an anomalous $\mathbb{Z}_{2a}\times\mathbb{Z}_{2b}$ symmetry given by (\ref{sigmatausymm}) and an ordinary (non-anomalous) $\mathbb{Z}_4$ spin chain (this was also noted in Ref.~\onlinecite{arkya2022}). 

\textbf{\emph{$\mathbb{Z}_4$ spin chain.}}---The $\mathbb{Z}_4$ spin chain is a spin chain where each spin can be in four different states. The two basic operators acting on the $j$th spin are the ``clock'' operator $C_j$ and the ``shift'' operator $S_j$. These operators take the following form (in the clock eigenstate basis):
\begin{equation}
C_j=\begin{pmatrix} 1 & 0 & 0 & 0\\0 & i & 0 & 0\\0 & 0 & -1& 0\\0 & 0 & 0& -i\end{pmatrix}\qquad S_j=\begin{pmatrix} 0 & 0 & 0 & 1\\1 & 0 & 0 & 0\\0 & 1 & 0& 0\\0 & 0 & 1& 0\end{pmatrix}.
\end{equation}
Note that $C_j$ and $S_j$ satisfy the algebra
\begin{equation}\label{clockalgebra}
C_j^4=S_j^4=\mathbbm{1}\qquad C_jS_j=iS_jC_j.
\end{equation}

In this paper, we will be interested in $\mathbb{Z}_4$ spin chains with a global $\mathbb{Z}_4$ symmetry given by $U_{\mathbb{Z}_4} = \prod_j S_j$. Such spin chains are closely related to the $\mathbb{Z}_2 \times \mathbb{Z}_2$ SPT edge theory described above. To see the relation, consider a symmetry breaking Hamiltonian of the form $H = -\sum_j \frac{1}{2}(C_j^\dagger C_{j+1} + C_{j+1}^\dagger C_j)$.  
This system has four degenerate ground states, and likewise four different species of domain walls. The different types of domain walls can be conveniently labeled by fourth roots of unity, $\{1, i, -1, -i\}$; the label associated with each domain wall is given by $C_j^\dagger C_{j+1}$ (assuming the domain wall is located between spins at sites $j$ and $j+1$). The crucial point is that these domain walls obey $\mathbb{Z}_4$ fusion rules just like the domain walls for the $\mathbb{Z}_2 \times \mathbb{Z}_2$ SPT edge theory, suggesting that there may be a way to map one system onto the other.


\textbf{\emph{Mapping between the models.}}---We will now map the Hilbert space of the $\mathbb{Z}_2 \times \mathbb{Z}_2$ SPT edge theory onto the Hilbert space of the $\mathbb{Z}_4$ spin chain. 

As we mentioned earlier, the basic idea is to map the four kinds of domain walls in the $\mathbb{Z}_4$ spin chain onto the four kinds of domain walls in the SPT edge theory. To do this, we need to map the spin chain operator $C_j^\dagger C_{j+1}$ (which measures $\mathbb{Z}_4$ domain walls) onto a corresponding domain wall operator in the SPT edge theory. The latter operator should have the four domain wall configurations in Fig.~\ref{fig:fig2} as eigenstates, with eigenvalues $1,i,-1,$ and $-i$. It should also be invariant under the $\mathbb{Z}_2 \times \mathbb{Z}_2$ symmetry, since we want our mapping to map $\mathbb{Z}_4$ symmetric operators in the spin chain (like $C_j^\dagger C_{j+1}$) onto $\mathbb{Z}_2 \times \mathbb{Z}_2$ symmetric operators in the SPT edge theory. These requirements are satisfied by the operator $\tau^x_{j+1/2}i^{(1-\sigma^z_j\sigma^z_{j+1})/2}$, so we map
\begin{equation}\label{cdef}
C_j^{\dagger}C_{j+1}\leftrightarrow \tau^x_{j+1/2}i^{(1-\sigma^z_j\sigma^z_{j+1})/2}.
\end{equation}
\begin{figure}[tb]
   \centering
   \includegraphics[width=.9\columnwidth]{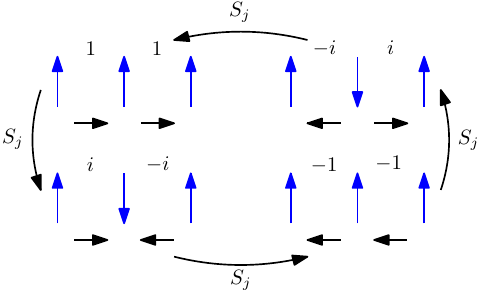} 
   \caption{The action of $S_j$ on domain wall states in the SPT edge theory: $S_j$ shifts the domain wall measured by $C_{j-1}^\dagger C_j$ by $i$ and the domain wall measured by $C_j^\dagger C_{j+1}$ by $-i$. Here, $j$ labels the spin in the middle of each 5-spin configuration.}
   \label{fig:fig3}
\end{figure}

In addition to $C_j^\dagger C_{j+1}$, we also need to work out how our mapping acts on the shift operator $S_j$. To do this, notice that $S_j$ shifts the domain wall measured by $C_{j-1}^\dagger C_j$ by $i$ and the domain wall measured by $C_j^\dagger C_{j+1}$ by $-i$. This means that $S_j$ should map to an operator whose action on SPT domain wall states is of the form shown in Fig.~\ref{fig:fig3}. 
Another requirement is that $S_j$ should map onto an operator that is invariant under the $\mathbb{Z}_2 \times \mathbb{Z}_2$ symmetry. These two requirements lead us to the mapping
\begin{align}
\begin{split}\label{sdef}
S_j&\leftrightarrow\sigma^x_j\left[\left(\frac{1+\sigma^z_{j-1}\sigma^z_j}{2}\right)+\left(\frac{1-\sigma^z_{j-1}\sigma^z_j}{2}\right)\tau^z_{j-1/2}\right]\\
&\times\left[\left(\frac{1+\sigma^z_j\sigma^z_{j+1}}{2}\right)\tau^z_{j+1/2}+\left(\frac{1-\sigma^z_j\sigma^z_{j+1}}{2}\right)\right].
\end{split}
\end{align}

Eqs.~(\ref{cdef}-\ref{sdef}) define a mapping between local $\mathbb{Z}_4$ symmetric operators in the spin chain and local $\mathbb{Z}_2 \times \mathbb{Z}_2$ symmetric operators in the SPT edge theory. To understand the \emph{global} properties of this mapping, we note that straightforward algebra shows that
\begin{equation}\label{symmetrymap}
\prod_jS_j \leftrightarrow U_a\qquad \prod_jC_j^\dagger C_{j+1} \leftrightarrow U_b.
\end{equation}

These equations tell us how the various symmetry sectors map onto one another. In particular, we see that the two sectors of the SPT edge theory with $U_a = \pm 1$ and $U_b = 1$ map onto the two
spin chain sectors with $\prod_j S_j = \pm 1$, and with \emph{periodic} boundary conditions. On the other hand, the two SPT sectors with $U_a=\pm1$ and $U_b=-1$ map onto spin chain sectors with $\prod_jS_j=\pm1$ and \emph{anti-periodic} boundary conditions. Here, the anti-periodic boundary condition can be implemented, on a closed loop of $N$ sites, by using $C_{N+1}=-C_1$ instead of $C_{N+1}=C_1$ (which corresponds to the periodic case). Putting this all together, we see that the Hilbert space of the SPT edge theory maps onto the Hilbert space of the $\mathbb{Z}_4$ spin chain with a particular combination of sectors, namely the two symmetry sectors $\prod_jS_j=\pm1$, with either periodic or anti-periodic boundary conditions. 

Alternatively, one can think of this particular combination of sectors as describing a $\mathbb{Z}_4$ spin chain coupled to a $\mathbb{Z}_2$ gauge field $\{\nu_{j+1/2}\}$ with the gauge constraint $\nu^x_{j-1/2}\nu^x_{j+1/2}=S_j^2$. In the gauged spin chain, the two boundary conditions correspond to sectors with even and odd $\mathbb{Z}_2$ gauge flux, while the global constraint $\prod_jS_j=\pm1$ is imposed by gauge invariance. In this paper, we will mostly work with the explicit sector description rather than the gauged spin chain language, but the latter is a completely equivalent way to think about our mapping.


\textbf{\emph{Using the mapping.}}---We will now use the mapping to understand the phases and phase transitions of the SPT edge theory. We start with the $\mathbb{Z}_4$ spin chain, which is expected to support three different gapped phases: an ordered phase where the $\mathbb{Z}_4$ symmetry is spontaneously broken, a disordered phase where the symmetry is unbroken, and a partially ordered phase where the $\mathbb{Z}_4$ symmetry is broken down to $\mathbb{Z}_2$\cite{jorge1977}. 
We can diagnose each of these phases in terms of an order parameter $O$, and a disorder parameter $D$ defined as follows\footnote{By some conventions, ``order parameter" refers to the operator $C_j$. Here, we mean the two-point correlation function in (\ref{order}), which is nonzero in the ordered phase and zero in the disordered phase.}: 
\begin{align}\label{order}
    O = \lim_{|i-j|\to\infty}\langle C_i^\dagger C_j\rangle
\quad
D = \lim_{|i-j|\to\infty}\left\langle \prod_{k=i}^jS_k\right\rangle.
\end{align}
Each phase has a different pattern of order and disorder parameters:
\begin{align}
    \text{Ordered phase}: \quad &O  \neq 0, \quad D = 0 \nonumber \\
    \text{Disordered phase}: \quad &O  = 0, \quad D \neq 0 \nonumber \\
    \text{Partially ordered phase}: \quad &O = 0, \quad D = 0.
\end{align}
Now, according to (\ref{symmetrymap}), our mapping takes the order parameter $O$ for the $\mathbb{Z}_4$ spin chain onto the symmetry transformation $U_b$ restricted to an interval, which is, by definition, a disorder parameter for $\mathbb{Z}_{2b}$. Likewise, our mapping takes the $\mathbb{Z}_4$ disorder parameter $D$ to a $\mathbb{Z}_{2a}$ disorder parameter. It follows that the ordered phase of the spin chain corresponds to a phase of the SPT edge theory with a vanishing $\mathbb{Z}_{2a}$ disorder parameter and a nonvanishing $\mathbb{Z}_{2b}$ disorder parameter -- i.e. a phase with broken $\mathbb{Z}_{2a}$ symmetry and unbroken $\mathbb{Z}_{2b}$ symmetry. By the same reasoning, the disordered phase of the spin chain maps onto a phase with unbroken $\mathbb{Z}_{2a}$ symmetry and broken $\mathbb{Z}_{2b}$ symmetry. Finally, the partially ordered phase of the spin chain maps onto a phase where both $\mathbb{Z}_{2a}$ and $\mathbb{Z}_{2b}$ are broken.


The most important application of these results, for our purposes, involves phase \emph{transitions}. In particular, consider a hypothetical critical point between the $\mathbb{Z}_{2a}$ broken ($\mathbb{Z}_{2b}$ unbroken) phase, and its partner, the $\mathbb{Z}_{2b}$ broken ($\mathbb{Z}_{2a}$ unbroken) phase. Applying our mapping, such critical points correspond to critical points between the ordered and disordered phase of the $\mathbb{Z}_4$ spin chain. This means that the problem of understanding DQCPs in the context of the SPT edge theory maps onto the problem of understanding ordinary symmetry breaking critical points for the $\mathbb{Z}_4$ spin chain. Since the latter critical points are known to exist and are well-understood, this proves the existence of DQCPs and also allows us deduce their structure.

\textbf{\emph{Exactly solvable model.}}---More concretely, we can use our mapping to construct an exactly solvable Hamiltonian that describes a continuous phase transition between the $\mathbb{Z}_{2a}$ broken ($\mathbb{Z}_{2b}$ unbroken) phase, and the $\mathbb{Z}_{2b}$ broken ($\mathbb{Z}_{2a}$ unbroken) phase of the SPT edge theory and therefore describes a DQCP. To build such a Hamiltonian, we start with an exactly solvable spin chain Hamiltonian that describes a $\mathbb{Z}_4$ symmetry breaking transition. In particular, we use the $\mathbb{Z}_4$ clock model:
\begin{align}
\begin{split}\label{hclock}
H_{\mathrm{clock}}(\alpha)&=-(1-\alpha)\sum_j \frac{1}{2}\left(C_j^\dagger C_{j+1}+C_{j+1}^\dagger C_j\right)\\
&-\alpha\sum_j\frac{1}{2}\left(S_j+S_j^\dagger\right).
\end{split}
\end{align}
Later, we will review how to solve $H_{\mathrm{clock}}$ exactly; for now, the only property we need is that $H_{\mathrm{clock}}$ belongs to the $\mathbb{Z}_4$ ordered phase for $\alpha<\frac{1}{2}$ and the disordered phase for $\alpha>\frac{1}{2}$, with a direct transition at $\alpha=\frac{1}{2}$.

To apply our mapping, we write
\begin{align}
H_{\mathrm{clock}}(\alpha) = (1-\alpha)H_{\mathrm{a, clock}} +
\alpha H_{b, \mathrm{clock}},
\end{align}
where $H_{\mathrm{a, clock}}$ and $H_{\mathrm{b, clock}}$ describe the two sets of terms in (\ref{hclock}). Notice that $H_{\mathrm{a, clock}}$ and $H_{b, \mathrm{clock}}$ are both sums of commuting terms. Furthermore, one can see that $H_{\mathrm{a, clock}}$ and $H_{b, \mathrm{clock}}$ belong to the ordered and disordered phases, respectively. Hence, applying our mapping to $H_{\mathrm{a, clock}}$ gives a commuting Hamiltonian describing the $\mathbb{Z}_{2a}$ broken ($\mathbb{Z}_{2b}$ unbroken) phase of the SPT edge theory:\footnote{Note that this Hamiltonian is not exactly the same as the Hamiltonian written under Eq.~\ref{sigmatausymm}, whose ground states also spontaneously break $U_a$.}
\begin{equation}\label{ha}
H_a=-\sum_j\left(\frac{1+\sigma^z_{j}\sigma^z_{j+1}}{2}\right)\tau^x_{j+1/2}.
\end{equation}

Similarly, applying our mapping to $H_{b, \mathrm{clock}}$, gives a commuting Hamiltonian for the $\mathbb{Z}_{2b}$ broken ($\mathbb{Z}_{2a}$ unbroken) phase:
\begin{align}
\begin{split}\label{hb}
H_b&=-\sum_j \Bigg[\sigma^x_j\left(\frac{1+\sigma^z_{j-1}\sigma^z_{j+1}}{2}\right)\left(\frac{\tau^z_{j-1/2}+\tau^z_{j+1/2}}{2}\right)\\
&+\sigma^x_j\left(\frac{1-\sigma^z_{j-1}\sigma^z_{j+1}}{2}\right)\left(\frac{1+\tau^z_{j-1/2}\tau^z_{j+1/2}}{2}\right)\Bigg].
\end{split}
\end{align}

Our exactly solvable model that tunes between these two symmetry breaking phases is given by
\begin{equation}\label{halpha}
H(\alpha)=(1-\alpha)H_a+\alpha H_b.
\end{equation}
Like $H_{\mathrm{clock}}$, this Hamiltonian describes a direct transition between the two phases (and hence a DQCP) at $\alpha=\frac{1}{2}$.

\textbf{\emph{Exactly solvable critical point.}}---We now review how to solve the $\mathbb{Z}_4$ clock model $H_{\mathrm{clock}}(\alpha)$ (\ref{hclock}) and hence also $H(\alpha)$. The basic idea is to map $H_{\mathrm{clock}}$ onto two decoupled transverse field Ising models which undergo simultaneous symmetry breaking transitions. To do this, we map each four dimensional spin onto two spin-1/2 degrees of freedom, denoted by $\mu_j$ and $\rho_{j}$ (note that  $\mu_j$ and $\rho_{j}$ should not be confused with $\sigma_j$ and $\tau_{j+1/2}$). We then write
\begin{equation}\label{transf}
    C_j=\frac{e^{-i\pi/4}}{\sqrt{2}}\left(\mu^z_j+i\rho^z_{j}\right)
    \end{equation}
and
\begin{equation}\label{transf2}
    S_j=\mu^x_j\left(\frac{1+\mu^z_j\rho^z_{j}}{2}\right)+\rho^x_{j}\left(\frac{1-\mu^z_j\rho^z_{j}}{2}\right).
\end{equation}

Using (\ref{transf}) and (\ref{transf2}), we compute
\begin{align}
\begin{split}\label{clockising}
    C_j^\dagger C_{j+1}+C_{j+1}^\dagger C_j&=\mu^z_j\mu^z_{j+1}+\rho^z_{j}\rho^z_{j+1}\\
    S_j+S_j^\dagger&=\mu^x_j+\rho^x_{j}.
\end{split}
\end{align}

Applying this map to the $\mathbb{Z}_4$ clock model in Eq.~\ref{hclock} gives
\begin{align}
\begin{split}
    H_{\mathrm{clock}}&=-(1-\alpha)\sum_j \frac{1}{2}\left(\mu^z_j\mu^z_{j+1}+\rho^z_{j}\rho^z_{j+1}\right)\\
&-\alpha\sum_j\frac{1}{2}\left(\mu^x_j+\rho^x_{j}\right),
\end{split}
\end{align}
which recovers the well-known fact that the $\mathbb{Z}_4$ clock model is unitarily equivalent to two decoupled transverse field Ising models. 

This mapping implies that the DQCP that occurs at $\alpha = 1/2$ is equivalent to two copies of the critical Ising theory\cite{dijkgraaf1989}. More precisely, the DQCP is equivalent to a particular combination of \emph{sectors} of the Ising theory: translating the sectors $\prod_j S_j = \pm 1$ and $C_{N+1} = \pm C_1$ into the Ising language, we see that $H(\alpha)$ is described by the symmetry sector $\prod_j \mu_j^x \rho_j^x = 1$, with the same (periodic or anti-periodic) boundary conditions in both $\mu, \rho$, i.e. $\mu^z_{N+1} = \pm \mu^z_1$ and $\rho^z_{N+1} = \pm \rho^z_1$.\footnote{To see that $\prod_jS_j=\pm1$ translates to $\prod_j \mu_j^x \rho_j^x = 1$, notice that $S_j^2=\mu^x_j\rho^x_j$ according to (\ref{transf2}). Since $\prod_jS_j=\pm1$, $\prod_jS_j^2=\prod_j\mu^x_j\rho^x_j=1$. The fact that the two Ising models must have the same boundary condition follows directly from $C_{N+1}=\pm C_1$, using the definition of $C_j$ in (\ref{transf}).}

Using this mapping we can obtain all the critical exponents of the DQCP. For example, the correlation length $\xi$ near the critical point diverges as $\xi \sim \frac{1}{|\alpha-\frac{1}{2}|^{\nu}}$ with $\nu = 1$. Also, the two-point correlators for the $\mathbb{Z}_{2a}$ and $\mathbb{Z}_{2b}$ order parameters $\sigma^z$ and $\tau^z$ are given by
\begin{equation}\label{critexp}
    \langle\sigma^z_i\sigma^z_j\rangle\sim\frac{1}{|i-j|^{1/2}}\qquad\langle\tau^z_{i+1/2}\tau^z_{j+1/2}\rangle\sim\frac{1}{|i-j|^{1/2}}.
\end{equation}



Is the above DQCP stable to perturbations? The answer to this question depends on what additional symmetries we impose beyond $\mathbb{Z}_{2a} \times \mathbb{Z}_{2b}$. For example, suppose we impose both time reversal and parity symmetry. In this case, it is well known that the critical point of the $\mathbb{Z}_4$ clock model does not have any relevant symmetric operators beyond the tuning parameter $\alpha$, but it does have a marginal operator corresponding to $\lambda\sum_j(C_j^2C_{j+1}^2+S_j^2)$. Adding this operator moves the system along a critical line\cite{jorge1977,kohmoto1981,lech2002}. Therefore the DQCP that we described above is actually part of deconfined quantum critical \emph{line} with continuously varying exponents (see the supplemental material\cite{supp} for more details). On the other hand, if we don't impose additional symmetries, then the critical point has other relevant symmetric operators that drive the system into a gapless phase, destroying the direct transition. These are known as ``chiral perturbations," and are given by $\lambda_{\varphi}\sum_j\left(C_j^\dagger C_{j+1}e^{i\varphi}+h.c.\right)$ and $\lambda_{\vartheta}\sum_j\left(S_je^{i\vartheta}+h.c.\right)$\cite{schulz1983,huse1984,nyckees2022,whitsitt2018}. More generally, if we consider the whole critical line, there is a region (i.e.~a range of $\lambda$) where the chiral perturbations are irrelevant\cite{kohmoto1981,schulz1983,huse1984,nyckees2022,supp}. In this region, time reversal symmetry and parity symmetry are not required to stabilize the transition.

\textbf{\emph{Self-duality at criticality.}}---An interesting aspect of the above DQCP is that it is \emph{self-dual}: there is a duality transformation that maps the critical point to itself and exchanges the $\mathbb{Z}_{2a}$ and $\mathbb{Z}_{2b}$ order parameters in (\ref{critexp}). This self-duality is reminiscent of the self-duality that occurs in other DQCPs, such as in the XY antiferromagnet/VBS transition obtained from adding easy-plane spin anisotropy to the N\'{e}el/VBS theory\cite{senthil2004,senthil2004_2,wang2017}. 

The duality transformation -- denoted by $U_c$ -- is easiest to understand in terms of the $\mathbb{Z}_4$ spin chain variables: in this description, $U_c$ maps $C_j^\dagger C_{j+1}$ onto $S_{j+1}$ and maps $S_{j}$ onto $C_j^\dagger C_{j+1}$. This is similar to the Kramers-Wannier duality, but unlike standard Kramers-Wannier duality, $U_c$ is both (1) unitary and (2) locality preserving, in the sense that it maps local operators to local operators. These properties are due to the unusual sector structure in our $\mathbb{Z}_4$ spin chain, or equivalently the fact that the $\mathbb{Z}_4$ spin chain is coupled to a $\mathbb{Z}_2$ gauge field (see the supplemental material\cite{supp} for more details). One consequence of the unitarity and locality of $U_c$ is that $U_c$ can also be viewed as an ordinary symmetry, rather than a duality. 

\textbf{\emph{Discussion.}}---
As emphasized above, at the core of our construction is the mapping between the $\mathbb{Z}_2\times\mathbb{Z}_2$ SPT edge theory and the $\mathbb{Z}_{4}$ spin chain (\ref{cdef}-\ref{sdef}). This mapping can be readily generalized to any $\mathbb{Z}_{N_1}\times\mathbb{Z}_{N_2}$ SPT edge theory with a primitive\footnote{By ``primitive'' we mean that the topological invariant $e^{i \Theta_{12}}$ defined in Ref.~\cite{topologicalinvariants} is a primitive $d$th root of unity where $d = \text{gcd}(N_1, N_2)$.} mixed anomaly. Specifically, any edge theory of this kind can be mapped onto a $\mathbb{Z}_{N_1 N_2}$ spin chain in such a way that the Landau forbidden transition in the edge theory maps onto an ordinary symmetry breaking transition in the spin chain. 

Moving forward, it would be interesting to find examples of these mappings for other kinds of anomalies, such as ``type-III anomalies''\cite{wang2015spt,zaletel2014}, or for non-Abelian symmetry groups. Examples of this kind could give solvable DQCPs with richer structure. It would also be interesting to generalize to higher dimensional systems, though this is not straightforward since our construction relies on charges and domain walls having the same dimensionality, as shown in Fig.~\ref{fig:fig2}.



Another interesting generalization is to add disorder to our model, by drawing the coefficients of the terms in $H_{a,\mathrm{clock}}$ and $H_{b,\mathrm{clock}}$ from random distributions. It was shown in Refs.~\onlinecite{fisher1,fisher2,senthilclock} that strongly disordered $\mathbb{Z}_N$ clock models have continuous transitions with critical properties that can be obtained exactly using a renormalization group analysis. In the corresponding SPT edge theory, this kind of model would give an example of a \emph{disordered} DQCP. 

C.Z. and M.L. thank T. Senthil for helpful discussions, and especially for suggesting disordered generalizations of our model, as mentioned in the discussion. C.Z. and M.L. also thank J. Wang for discussions related to type-III anomalies. C.Z. and M.L. acknowledge the support of the Kadanoff Center for Theoretical Physics at the University of Chicago, the Simons Collaboration on Ultra-Quantum Matter, which is a grant from the 
Simons Foundation (651440, M.L.), and the National Science Foundation Graduate Research Fellowship under Grant No. 1746045. 
\bibliography{dqcpbib}

\end{document}


\title{Supplemental Material}

\author{Carolyn Zhang}
\author{Michael Levin}
\affiliation{Department of Physics, Kadanoff Center for Theoretical Physics, University of Chicago, Chicago, Illinois 60637,  USA}

\maketitle
\section{Field theory description of deconfined quantum critical line}
In this section, we present a field-theoretic description of the deconfined quantum critical line that separates the two symmetry breaking phases of the SPT edge theory. The exactly solvable DQCP presented in the main text corresponds to a particular point along this line, as we explain below.

We begin with a field theory description of the SPT edge theory. This field theory consists of two bosonic fields $\theta, \phi$ with the commutation relation
\begin{equation}\label{commutator}
\left[\theta(x'), \partial_x\phi(x)\right]=2\pi i\delta(x-x'),
\end{equation}
where the fundamental local operators in the SPT edge theory are $e^{\pm i \theta}, e^{\pm i \phi}$. The $\mathbb{Z}_{2a} \times \mathbb{Z}_{2b}$ symmetries act on 
$\theta$ and $\phi$ as follows:
\begin{align}
\begin{split}\label{symmaction}
&U_a:\phi\to\phi+\pi\qquad\theta\to\theta\\
&U_b:\phi\to\phi\qquad\theta\to\theta+\pi.
\end{split}
\end{align}
(To see that Eq.~(\ref{symmaction}) describes the desired mixed anomaly between $\mathbb{Z}_{2a}$ and $\mathbb{Z}_{2b}$, observe that the symmetry actions $U_a$ and $U_b$ are not anomalous as individual $\mathbb{Z}_2$ symmetries, while the product $U_b\cdot U_a$ \emph{is} anomalous\cite{levin2012,else2014}. This also holds for the lattice model in the main text and is characteristic of the mixed anomaly for $\mathbb{Z}_{2a} \times \mathbb{Z}_{2b}$). 


%

Next, observe that we can realize the $\mathbb{Z}_{2a}$ broken ($\mathbb{Z}_{2b}$ unbroken) phase of the SPT edge theory with the following edge Hamiltonian:
\begin{equation}
\mathcal{H}=\frac{1}{8\pi}\left[K\left(\partial_x\theta\right)^2+\frac{4}{K}\left(\partial_x\phi\right)^2\right]+A_{\phi}\cos(2\phi),
\end{equation}
Here, $K$ is the Luttinger parameter, $A_{\phi}$ is a real coefficient, and we work in units where the edge velocity $v$ is set to $v=1$. The above Hamiltonian breaks the $\mathbb{Z}_{2a}$ symmetry spontaneously in a regime where the term $A_{\phi}\cos(2\phi)$ is relevant. In this case, $\phi$ becomes fixed to one of the two minima of the cosine potential leading to an expectation value for $e^{i \phi}$, which is odd under $U_a$. Likewise, we can realize the $\mathbb{Z}_{2b}$ broken ($\mathbb{Z}_{2a}$ unbroken) phase using the edge Hamiltonian
\begin{equation}
\mathcal{H}=\frac{1}{8\pi}\left[K\left(\partial_x\theta\right)^2+\frac{4}{K}\left(\partial_x\phi\right)^2\right]+A_{\theta}\cos(2\theta),
\end{equation}
As before, if we are in regime where $A_{\theta}\cos(2\theta)$ is relevant, then $e^{i \theta}$ acquires an expectation value, spontaneously breaking the $\mathbb{Z}_{2b}$ symmetry.

\begin{figure}[tb]
   \centering
   \includegraphics[width=.9\columnwidth]{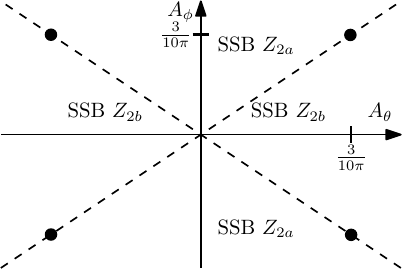} 
   \caption{Phase diagram for the Hamiltonian in (\ref{fieldtheory}) for $K=2$. The dotted lines are critical lines, with continuously varying exponents. The black dots at $(A_{\theta},A_{\phi})=\left(\pm \frac{3}{10\pi},\pm\frac{3}{10\pi}\right)$ indicate points along the critical lines described by the exactly solvable critical points from the main text.}
   \label{fig:phasediag}
\end{figure}

The above considerations suggest a natural field theory for the deconfined quantum critical line separating the two symmetry breaking phases: 
\begin{equation}\label{fieldtheory}
\mathcal{H}=\frac{1}{8\pi}\left[K\left(\partial_x\theta\right)^2+\frac{4}{K}\left(\partial_x\phi\right)^2\right]+A_{\theta}\cos(2\theta)+A_{\phi}\cos(2\phi),
\end{equation}
Only one of the two cosine operators is relevant for a given value of $K$: the scaling dimension of $\cos(2\phi)$ is $K$ while the scaling dimension of $\cos(2\theta)$ is $\frac{4}{K}$. Therefore, $\cos(2\phi)$ is relevant for $K<2$ and $\cos(2\theta)$ is relevant for $K>2$. At $K=2$, both cosines are marginal. For this value of $K$, the two marginal operators $\cos(2\phi)$ and $\cos(2\theta)$ compete, and the fate of the theory depends on the magnitudes of the coefficients $A_\phi$, $A_\theta$ (see Fig.~\ref{fig:phasediag}). In particular, if $|A_\phi| > |A_\theta|$ then the $\mathbb{Z}_{2a}$ symmetry is spontaneously broken, while if $|A_\phi| < |A_\theta|$ then the $\mathbb{Z}_{2b}$ symmetry is spontaneously broken. The deconfined quantum critical line(s) occur when $A_\phi = \pm A_\theta$.

A few comments about (\ref{fieldtheory}): first, we note that the field theory (\ref{fieldtheory}) has been used in various contexts in the literature (see e.g. Ref~\onlinecite{lech2002} and references therein), and is similar to the field theory of the 1D DQCP discussed in Ref.~\onlinecite{jiang2019}. We also note that if we define $\theta'=2\theta$ and $\phi'=\phi/2$, the resulting field theory for $\theta'$ and $\phi'$ is precisely the field theory that describes the $\mathbb{Z}_4$ spontaneous symmetry breaking critical line\cite{wen2004}. This mapping to the $\mathbb{Z}_4$ symmetry breaking transition is the field theory analog of the lattice mapping discussed in the main text.

We now review how to derive the properties of the critical lines with $A_\phi = \pm A_\theta$. To do this, we fix $K=2$ and reparameterize the two cosines in (\ref{fieldtheory}) in symmetric and anti-symmetric combinations:
\begin{align}
A_{\theta}\cos(2\theta)+A_{\phi}\cos(2\phi) &= 
g_+ [\cos(2\theta) + \cos(2 \phi)] \nonumber \\
&+ g_- [\cos(2\theta) - \cos(2 \phi)]
\end{align}
The critical lines occur when either $g_{+}=0$ or $g_{-}=0$. For concreteness, consider the critical line with $g_- = 0$. The key to analyzing this line is to utilize the hidden $SU(2)$ symmetry in this model. In particular, there exists an $SU(2)$ rotation that maps
\begin{align}
\cos(2\theta) + \cos(2 \phi) \rightarrow \frac{1}{2} [(\partial_x \theta)^2 - (\partial_x \phi)^2]
\end{align}
and leaves $\cos(2\theta) - \cos(2 \phi)$ and $(\partial_x \theta)^2 + (\partial_x \phi)^2$ invariant. After this (unitary) $SU(2)$ transformation, our Hamiltonian becomes
\begin{align} \label{Hgaussian}
\mathcal{H}& \rightarrow \frac{\tilde{v}}{8\pi}\left[\tilde{K}\left(\partial_x\theta\right)^2+\frac{4}{\tilde{K}}\left(\partial_x\phi\right)^2\right] \nonumber \\
&+ g_- [\cos(2\theta) - \cos(2 \phi)]
\end{align}
where
\begin{equation}\label{Ktilde}
    \tilde{K}^2=4\frac{1+2\pi g_{+}}{1-2\pi g_{+}}\qquad \tilde{v}^2=1-4\pi^2g_{+}^2.
\end{equation}
Crucially, along the critical line $g_- = 0$, the $SU(2)$ rotated Hamiltonian (\ref{Hgaussian}) is \emph{Gaussian}. As a result, critical exponents can be computed straightforwardly. These critical exponents vary continuously along the critical line due to the continuously varying Luttinger parameter $\tilde{K}$.



To complete this discussion, we now work out which point(s) along these critical lines correspond to the exactly solvable lattice model given in Eq.~15. To this end, recall that in the main text, we mapped our solvable lattice model onto two decoupled Ising models. This mapping implies that the operator that drives the transition is the sum of the energy fields of the two Ising models, which has a scaling dimension $1$. In comparison, in the field theory (\ref{Hgaussian}), the operator that drives the transition is $[\cos(2\theta) - \cos(2\phi)]$ which has a scaling dimension of $\text{min}(\frac{4}{\tilde{K}}, \tilde{K})$ (since the scaling dimensions of $\cos(2\theta)$ and $\cos(2\phi)$ are $\frac{4}{\tilde{K}}$ and $\tilde{K}$, respectively). Setting
$\frac{4}{\tilde{K}}=1$ and $\tilde{K}=1$, we see that our exactly solvable lattice model corresponds to the two points
$g_+ = \pm\frac{3}{10\pi}$ for the critical line with $g_- = 0$. Similarly, along the other critical line with $g_+ = 0$, the solvable model corresponds to the two points with $g_- = \pm\frac{3}{10\pi}$. These points are marked in the phase diagram shown in Fig.~\ref{fig:phasediag}.


\section{Chiral perturbations}
In the main text, we mentioned that, if we don't impose time reversal and parity symmetry, then there exist chiral perturbations that can destroy our exactly solvable DQCP and drive the system into a gapless phase. In this section, we analyze these chiral perturbations using the field theory (\ref{fieldtheory}) and we identify the regions of the critical lines where these perturbations are relevant/irrelevant. In the regions where they are relevant (such as in our exactly solvable lattice model), we need time reversal and parity symmetry to stabilize the direct transition, while in the regions where they are irrelevant, we do not need any additional symmetries.


In the field theory (\ref{fieldtheory}), the chiral perturbations correspond to $\partial_x\phi$ and $\partial_x\theta$. Note that these operators are invariant under $U_a$ and $U_b$, but they break either time reversal or parity symmetry, if we define
\begin{align}
\begin{split}
    T&:\phi\to\phi\qquad \theta\to -\theta\\
    P&:\phi\to-\phi\qquad\theta\to\theta
\end{split}
\end{align}


To determine when the chiral perturbations are relevant and when they are not, we must perform the same $SU(2)$ transformation as we described in the previous section on these operators. It is easiest to first take linear combinations of the two chiral perturbations, given by $\partial_x(\phi+\theta)$ and $\partial_x(\phi-\theta)$. The $SU(2)$ rotation maps $\partial_x(\phi+\theta)$ and $\partial_x(\phi-\theta)$ to $\cos(\phi+\theta)$ and $\cos(\phi-\theta)$ respectively. The scaling dimension of $\cos(\phi+\theta)$ and $\cos(\phi-\theta)$, which we call $\Delta_c$, is given by
\begin{equation}\label{dimc}
    \Delta_c=\frac{1}{\tilde{K}}+\frac{\tilde{K}}{4}
\end{equation}

The chiral perturbation is relevant when $\Delta_c<2$, which corresponds to
\begin{equation}\label{critk} (4-2\sqrt{3})<\tilde{K}<(4+2\sqrt{3})
\end{equation}


%
In this range, we must require time reversal symmetry and parity symmetry in order to exclude these perturbations and stabilize the direct transition. Outside of this range, the chiral perturbations are irrelevant, so we do not need to impose these extra symmetries.

\section{Duality transformation}
In this section, we explain how the duality transformation $U_c$ (which is also a unitary symmetry) acts on the different operators in the SPT edge theory. Our strategy will be to first work out how $U_c$ acts on the operators in the $\mathbb{Z}_4$ spin chain; we will then translate this over to the SPT edge theory. 

To understand how $U_c$ acts on the $\mathbb{Z}_4$ spin chain variables, it is convenient to use the gauge theory language where we think of the $\mathbb{Z}_4$ spin chain as coupled to a $\mathbb{Z}_2$ gauge field. As we mentioned in the main text, the gauge theory language is an alternative way to describe the sector structure of the $\mathbb{Z}_4$ spin chain. To use the gauge theory language, we introduce a $\mathbb{Z}_2$ gauge field $\{\nu_{j+1/2}\}$ on the links of the lattice, and we impose the gauge constraint $\nu^x_{j-1/2}\nu^x_{j+1/2}=S_j^2$. The local gauge invariant operators in the $\mathbb{Z}_4$ spin chain are
$\{C_j^\dagger \nu^z_{j+1/2}C_{j+1}, S_j, C_j^2, \nu^x_{j+1/2}\}$. All other gauge invariant operators can be built from these operators, so it suffices to understand how $U_c$ acts on these operators.

The action of the duality transformation $U_c$ on the first two operators, $\{C_j^\dagger \nu^z_{j+1/2}C_{j+1}, S_j\}$ is similar to standard Kramers-Wannier duality:
\begin{align}
\begin{split}\label{symmclock1}
    U_c^\dagger C_j^\dagger\nu^z_{j+1/2}C_{j+1}U_c&=S_{j+1}\\
    U_c^\dagger S_jU_c&=C_{j}^\dagger\nu^z_{j+1/2}C_{j+1}.
\end{split}
\end{align}

To figure out the action of $U_c$ on the other two gauge invariant operators,
$\{C_j^2, \nu^x_{j+1/2}\}$, we use the Gauss law constraint $\nu^x_{j-1/2}\nu^x_{j+1/2}=S_{j}^2$ to deduce that
\begin{align}
\begin{split}\label{c2s2}
    U_c^\dagger C_j^2C_{j+1}^2U_c&=S_{j+1}^2=\nu^x_{j+1/2}\nu^x_{j+3/2}\\
    U_c^\dagger \nu^x_{j-1/2}\nu^x_{j+1/2}U_c &= U_c^\dagger S_j^2U_c =C_j^2C_{j+1}^2.
\end{split}
\end{align}

In order for $U_c$ to be locality preserving and satisfy (\ref{c2s2}), it must have the following action on $C_j^2$ and $\nu^x_{j+1/2}$: 
 \begin{align}
 \begin{split}\label{symmclock2}
    U_c^\dagger C_j^2U_c&=\nu^x_{j+1/2}\\
    U_c^\dagger \nu^x_{j+1/2}U_c&=C_{j+1}^2.
\end{split}
\end{align}
(As an aside, note that the above formulas imply that $U_c^2$ acts like a unit translation on the $\mathbb{Z}_4$ spin chain).

Now that we have figured out how $U_c$ acts on the operators in the (gauged) $\mathbb{Z}_4$ spin chain, we can translate this action to the SPT edge theory using our mapping. More specifically, we will need the mapping between the \emph{gauged} spin chain and the SPT edge theory. This mapping is essentially identical to the one described in Eqs.~(5-6) in the main text, but with the replacement $C_j^\dagger C_{j+1}\to C_j^\dagger \nu^z_{j+1/2}C_{j+1}$:
\begin{align}
\begin{split}\label{operatormap}
&C_j^{\dagger}\nu^z_{j+1/2}C_{j+1} \leftrightarrow \tau^x_{j+1/2}i^{(1-\sigma^z_j\sigma^z_{j+1})/2}\\
&S_j \leftrightarrow \sigma^x_j\left[\left(\frac{1+\sigma^z_{j-1}\sigma^z_j}{2}\right)+\left(\frac{1-\sigma^z_{j-1}\sigma^z_j}{2}\right)\tau^z_{j-1/2}\right]\\
&\times\left[\left(\frac{1+\sigma^z_j\sigma^z_{j+1}}{2}\right)\tau^z_{j+1/2}+\left(\frac{1-\sigma^z_j\sigma^z_{j+1}}{2}\right)\right]
\end{split}
\end{align}

We can also derive how the mapping acts on $C_j^2$ and $\nu^x_{j+1/2}$, by squaring the two equations in (\ref{operatormap}) and using the constraint $\nu^x_{j-1/2}\nu^x_{j+1/2}=S_{j}^2$:
\begin{align}
&C_j^2 \leftrightarrow \sigma^z_j, \qquad \nu^x_{j+1/2} \leftrightarrow \tau^z_{j+1/2}.
\label{operatormap2}
\end{align}



We are now ready to work out how the duality transformation $U_c$ acts on the operators in the SPT edge theory. It suffices to work out the action of $U_c$ on $\{\sigma^z_j, \tau^z_{j+1/2}, \sigma^x_j, \tau^x_{j+1/2}\}$ since all other operators can be built out of these. For the first two operators, $\{\sigma^z_j, \tau^z_{j+1/2}\}$, we can work out the action of $U_c$ using Eqs.~(\ref{symmclock1}), (\ref{symmclock2}) and (\ref{operatormap2}):
\begin{align}
\begin{split}\label{sigmaz}
    U_c^\dagger\sigma^z_{j}U_c&=\tau^z_{j+1/2}\\
    U_c^\dagger \tau^z_{j+1/2}U_c&=\sigma^z_{j+1}.
\end{split}
\end{align}
Thus, $U_c$ exchanges the $\mathbb{Z}_{2a}$ and $\mathbb{Z}_{2b}$ order parameters -- as we mentioned in the main text.

The duality transformation $U_c$ has a more complicated action on $\sigma^x_{j}$ and $\tau^x_{j+1/2}$. To obtain the action of $U_c$ on $\sigma^x_j$, we first solve for $\sigma^x_j$ in terms of gauged $\mathbb{Z}_4$ spin chain operators. From (\ref{operatormap}), we have
\begin{align}
\begin{split}
    \sigma^x_j&\leftrightarrow \left[\left(\frac{1+\sigma^z_{j-1}\sigma^z_j}{2}\right)+\left(\frac{1-\sigma^z_{j-1}\sigma^z_j}{2}\right)\tau^z_{j-1/2}\right]\\
&\times\left[\left(\frac{1+\sigma^z_j\sigma^z_{j+1}}{2}\right)\tau^z_{j+1/2}+\left(\frac{1-\sigma^z_j\sigma^z_{j+1}}{2}\right)\right]S_j^\dagger.
\end{split}
\end{align}

Writing the right hand side in terms of gauged spin chain operators, we get
\begin{align}
\begin{split}
\sigma^x_j&\leftrightarrow\left[\left(\frac{1+C_{j-1}^2C_j^2}{2}\right)+\left(\frac{1-C_{j-1}^2C_j^2}{2}\right)\nu^x_{j-1/2}\right]\\
&\times\left[\left(\frac{1+C_j^2C_{j+1}^2}{2}\right)\nu^x_{j+1/2}+\left(\frac{1-C_j^2C_{j+1}^2}{2}\right)\right]S_j^\dagger.
\end{split}
\end{align}

Then using the action of $U_c$ on the gauged spin chain operators (\ref{symmclock1}-\ref{c2s2}) and translating back into the SPT edge theory degrees of freedom using (\ref{operatormap}-\ref{operatormap2}), we get
\begin{align}
\begin{split}\label{sigmax}
    U_c^\dagger\sigma^x_jU_c&=\left[\left(\frac{1+\tau^z_{j-1/2}\tau^z_{j+1/2}}{2}\right)+\left(\frac{1-\tau^z_{j-1/2}\tau^z_{j+1/2}}{2}\right)\sigma^z_j\right]\\
&\times\left[\left(\frac{1+\tau^z_{j+1/2}\tau^z_{j+3/2}}{2}\right)\sigma^z_{j+1}+\left(\frac{1-\tau^z_{j+1/2}\tau^z_{j+3/2}}{2}\right)\right]\\
&\times (-i)^{(1-\sigma^z_j\sigma^z_{j+1})/2}\tau^x_{j+1/2}.
\end{split}
\end{align}

Similarly, solving for $\tau^x_{j+1/2}$, gives 
\begin{align}
\begin{split}
    \tau^x_{j+1/2}&=i^{(1-\sigma^z_j\sigma^z_{j+1})/2}C_{j+1}^\dagger\nu^z_{j+1/2}C_{j}\\
    &=i^{(1-C_j^2C_{j+1}^2)/2}C_{j+1}^\dagger\nu^z_{j+1/2}C_{j}.
\end{split}
\end{align}

Again using the transformation laws for the spin chiain (\ref{symmclock1}-\ref{c2s2}) and translating back into the SPT edge theory degrees of freedom using (\ref{operatormap}-\ref{operatormap2}), we obtain
\begin{align}
\begin{split}\label{taux}
    U_c^\dagger\tau^x_{j+1/2}U_c
    &=i^{(1-\tau^z_{j+1/2}\tau^z_{j+3/2})/2}\\
    &\times\left[\left(\frac{1+\sigma^z_{j}\sigma^z_{j+1}}{2}\right)+\left(\frac{1-\sigma^z_{j}\sigma^z_{j+1}}{2}\right)\tau^z_{j+1/2}\right]\\
&\times\left[\left(\frac{1+\sigma^z_{j+1}\sigma^z_{j+2}}{2}\right)\tau^z_{j+3/2}+\left(\frac{1-\sigma^z_{j+1}\sigma^z_{j+2}}{2}\right)\right]\sigma^x_{j+1}.
\end{split}
\end{align}

Eqs.~(\ref{sigmaz}), (\ref{sigmax}), and (\ref{taux}) completely define the action of $U_c$ on all operators in the SPT edge theory.

\bibliography{dqcpbib}